# Head-and-neck multi-channel B1$^+$ mapping and carotid arteries RF shimming using a parallel transmit head coil

Matthijs H.S. de Buck[1], Peter Jezzard[1], and Aaron T. Hess[1]

[1]Wellcome Centre for Integrative Neuroimaging, FMRIB Division, Nuffield Department of Clinical Neurosciences, University of Oxford, Oxford, United Kingdom

Word Count: 4696

Figures: 10

**Correspondence address**:

Aaron T. Hess, PhD

FMRIB Division, Wellcome Centre for Integrative Neuroimaging

University of Oxford

John Radcliffe Hospital

Oxford, OX3 9DU

United Kingdom

Email: aaron.hess@ndcn.ox.ac.uk

# Abstract

**Purpose:** Neurovascular MRI suffers from a rapid drop in $B_1^+$ into the neck when using transmit head coils at 7T. One solution to improving $B_1^+$ magnitude in the major feeding arteries in the neck is to use custom RF shims on parallel transmit (pTx) head coils. However, calculating such shims requires robust multi-channel $B_1^+$ maps in both the head and the neck, which is challenging due to low RF penetration into the neck, limited dynamic range of multi-channel $B_1^+$ mapping techniques, and B0 sensitivity. We therefore sought a robust large-dynamic-range pTx field mapping protocol, and tested whether RF shimming can improve carotid artery $B_1^+$ in practice.

**Methods:** A pipeline is presented that combines $B_1^+$ mapping data acquired using circularly polarized (CP-) and CP2-mode RF shims at multiple voltages. The pipeline was evaluated by comparing the predicted and measured $B_1^+$ for multiple random transmit shims, and by assessing the ability of RF shimming to increase the $B_1^+$ in the carotid arteries.

**Results:** The proposed method achieved good agreement between predicted and measured $B_1^+$ in both the head and the neck. The $B_1^+$ magnitude in the carotid arteries can be increased by 42% using tailored RF shims or by 37% using universal RF shims, while also improving the RF homogeneity compared to CP mode.

**Conclusion:** $B_1^+$ in the neck can be increased using RF shims calculated from multi-channel $B_1^+$ maps in both the head and the neck. This can be achieved using universal phase-only RF shims, facilitating easy implementation in existing sequences.

# 1. Introduction

Ultra-high-field MRI offers an increased signal-to-noise ratio (SNR) and longer T1-relaxation time for both tissue and blood. At 7T these properties have the potential to improve the contrast, resolution, and imaging time of intracranial neurovascular modalities such as cerebral angiography and perfusion, as well as the visualization of vessel wall pathology. However, increasing main magnetic field strength also introduces limitations due to increased specific absorption rate (SAR) and reduced homogeneity and spatial extent of the transmit magnetic field[1] ($B_1^+$).

When using typical transmit head coils at 7T, intracranial neurovascular imaging methods such as arterial spin labelling (ASL)[2,3] and intracranial vessel wall imaging[4] suffer from the rapid drop in $B_1^+$ into the neck. For ASL, this drop in $B_1^+$ limits the ability to effectively invert the inflowing blood in upstream labelling planes, such as in the carotid arteries in the neck. For vessel wall imaging, it reduces the ability to suppress the signal in upstream arterial blood, which is required to provide sufficient black-blood contrast between the vessel wall and the inflowing arterial blood. Although higher nominal flip angles can be applied to increase the inversion or saturation efficiency of arterial blood in low-$B_1^+$ areas, this is in practice constrained by a quadratic increase in SAR and by adverse effects on the magnetization of stationary spins within higher $B_1^+$ imaging regions. Dielectric pads[5,6] can be positioned near the the neck to increase both the transmit and receive sensitivity[7]. However, the use of dielectric pads increases experimental complexity and does not provide the ability to change the $B_1^+$ field over time, for example between different acquisitions or between signal preparation and readout modules within a single acquisition.

Improved control over $B_1^+$ in the neck can also be achieved using parallel transmission (pTx)[8] coils, which consist of multiple separate transmit channels. pTx provides improved control over the $B_1^+$ field by manipulating the amplitude and/or phase of each of the individual transmit channels. This can be used to achieve improved spatial homogeneity of the $B_1^+$ field[9], to reduce the SAR, to achieve spatial[10,11] and spectral[12] selectivity, or to increase the $B_1^+$ magnitude within a particular region of interest. Therefore, the use of pTx coils for neurovascular imaging could be used to improve the $B_1^+$ in the feeding arteries in the neck, thereby allowing improved inversion or saturation of inflowing arterial blood.

Most conventional head pTx coils are designed for imaging the brain, therefore consisting of transmit elements located near the brain. To improve the $B_1^+$ coverage in the brainstem, the cerebellum, and the carotid arteries, previous work proposed custom pTx coil designs that consist of transmit elements surrounding both the brain and the neck (using fixed[13] or geometrically adjustable[14] transmit arrays). However, the use of such coil designs adds experimental complexity and expense relative to the use of conventional pTx coil designs. Therefore, this work focuses on the potential of improving $B_1^+$ in the neck using conventional head pTx coils.

Before it is possible to calculate and optimize the achieved $B_1^+$ field using pTx methods, the transmit field of all individual pTx channels has to be characterized. Such multi-channel $B_1^+$ mapping[15] aims to measure the transmitted magnitude and relative phase of each transmit channel. Acquiring these data in both the head and the neck using pTx head coils can be challenging due to a combination of low radiofrequency (RF) penetration into the neck and the inherently limited dynamic range[15] of $B_1^+$ mapping techniques. Utilizing $B_1^+$ mapping techniques with typical transmit voltages provides accurate data in the brain, but does not provide useful information in the low-$B_1^+$ areas in the neck. Conversely, using high transmit voltages can achieve improved $B_1^+$ coverage in the neck, but is inaccurate for high-$B_1^+$ regions in the head. Therefore, this paper proposes and validates an approach which combines $B_1^+$ mapping data acquired at 7T using two complementary RF shims (to ensure adequate overall coverage) and at multiple transmit voltages to allow robust reconstruction of multi-channel $B_1^+$ maps for both the head and the neck.

Subsequently, multi-channel B1$^+$ data acquired using the proposed method are used to investigate by what means and by how much RF shimming[9] can improve the B1$^+$ magnitude in the major feeding arteries in the neck region. The transmit field strength in circular polarization (CP) mode and the potential increase along the carotid arteries using RF shimming are first discussed, followed by an investigation into the trade-off between improving B1$^+$ magnitude and B1$^+$ homogeneity in the arteries in the neck. Finally, to assess the feasibility of implementing RF shims for the carotid arteries in a clinical workflow without requiring time-consuming tailored shims, the possibility of using universal[16] RF shims is investigated, and the minimum number of subjects required for generating such universal RF shims for the arteries in the neck is evaluated.

# 2. Methods

### 2.1 Wide dynamic range multi-channel B1$^+$ mapping

In order to obtain robust multi-channel B1$^+$ maps, a three-step process was used, as summarized in Figure 1. First, absolute B1$^+$ maps were acquired using two standard RF shims (CP-mode – namely 45° phase increments per channel, and CP2-mode – namely 90° phase increments per channel), each with a range of transmit voltages. This was used to ensure that all required regions could be covered, since CP and CP2 modes are in general complementary in their sensitivity patterns. Additionally, a B0 map was acquired to correct for static field inhomogeneity. These were combined to form a single absolute B1$^+$ map in Hz/V. Secondly, channel-by-channel relative B1$^+$ maps were acquired, again at multiple voltages. Thirdly, the relative maps were combined with the absolute B1$^+$ map to achieve consistent absolute multi-channel B1$^+$ maps in both the head and neck.

For the first part, six absolute maps were acquired using a sandwiched train pre-saturated TurboFLASH image acquisition approach[17,18]. These included three transmit reference voltages (50 V, 100 V, and 175 V), each acquired in CP-mode and CP2-mode. Scan parameters for each individual 3D acquisition include TR = 1 s, TD = 0 s (using the sandwich scheme[17]), TE = 1.78 ms, flip angle = 9°, preparation flip angle = 90° (using a 500 µs rectangular pulse), bandwidth = 489 Hz/px, and scan time = 36 seconds.

The B1$^+$ maps acquired at different transmit voltages were combined by imposing a series of consistency criteria to the B1$^+$ data expressed in voltage-independent Hz/V units. The criteria were designed to cover the full dynamic range of B1$^+$ without departing the linear response of the measurement method, enabling full head-and-neck coverage. For each voxel in each image, the measured value is excluded from the final (combined) B1$^+$ map if:

1. The reference image ($S_0$) signal magnitude is smaller than the noise standard deviation; or
2. For data acquired at 100 V and 175 V, if the measured B1$^+$ magnitude (in voltage-independent Hz/V units) is substantially (using a tolerance margin of 0.5 Hz/V) smaller than the value measured in the corresponding voxel in an acquisition using a lower transmit reference voltage; or
3. For data acquired at 50 V and 100 V, if the measured B1$^+$ magnitude is smaller than the value measured in the corresponding voxel in an acquisition using a higher transmit reference voltage.

Here, Criterion 2 aims to remove values which are outside the linear regime between the measured and nominal flip angle[18], and Criterion 3 assumes that higher voltage acquisitions are more likely to be reliable and less likely to be below the lower limit of the dynamic range of the measurement method. If only one transmit voltage value remained for a given voxel after applying all three exclusion criteria, that value was used for the combined B1$^+$ map. If multiple transmit voltages remained (due to being within the tolerance margin), the average value of those different values was used (Figure 2i). The resulting B1$^+$ maps and inclusion maps in an exemplar acquisition are shown in Figure 2. A single sandwiched train pre-saturated TurboFLASH acquisition using a transmit reference voltage of 60 V and using same RF coil as this study has previously reported a dynamic range of a factor of 4 (ranging from 40° to 120°)[17]. Therefore, using transmit voltages of 50 V, 100 V, and 175 V, the multi-voltage approach as proposed here is expected to provide accurate results over a dynamic range of $B_1^+$ of a factor of 10.5, since the highest voltage of 175 V will enable $B_1^+$ regions that are as low as 13.7° at 60 V to be characterized accurately (60 V / 175 V * 40° at the lower end of the linear range of the method), and the lowest voltage of 50 V will enable $B_1^+$ regions that are as high as 144° at 60 V to be characterized accurately (60 V / 50 V * 120° at the upper end of the linear region of the method).

This multi-voltage acquisition combination step for the absolute $B_1^+$ maps was followed by the application of B1TIAMO[19] to combine the data acquired using CP-mode and CP2-mode (Figure 3c-e). B1TIAMO combines $B_1^+$ maps acquired using different RF shims to provide increased coverage in areas with low $B_1^+$ for any given shim by taking a weighted average of the $B_1^+$ based on the signal levels of the respective reference images.

B0 maps were also collected (acquired using a 3D gradient-recalled echo acquisition with TR = 4.9 ms, $TE_1/TE_2$ = 1.02/3.06 ms, flip angle = 7°, receive bandwidth = 538 Hz/px, scan time = 1:39 minutes) and used to correct for off-resonance $B_1^+$ underestimation arising from the 500 μs rectangular RF pulse used for SatTFL $B_1^+$ mapping: see Figure 3e-f. Here, PRELUDE[20] was used to unwrap the phase maps used for B0 measurement (Figure 3b).

The second step in the pipeline measured relative transmit maps at the same three reference voltages (50 V, 100 V, and 175 V) using low flip angle gradient-echo acquisitions (TR = 2.90 ms, TE = 1.02 ms, nominal flip angle = 7°, bandwidth = 500 Hz/px, scan time = 30 seconds per single voltage acquisition). These were measured by transmitting on one channel at a time in an interleaved fashion to minimise magnetisation history effects. Transmit maps at the three voltages were combined using a previously described large dynamic range method[21].

The final step combined the absolute maps with the relative maps to form complex multi-channel $B_1^+$ field maps (bottom row in the pipeline in Figure 1).

### 2.2 In vivo experiments

Data were acquired in 10 healthy volunteers (23-56 years old; 8 male/2 female). All acquisitions ($B_1^+$, B0, and structural data) were performed in 3D using the same FOV (225×225×300 mm) in coil coordinates. MPRAGE structural data were acquired at 1.2 mm isotropic resolution for anatomical reference. For $B_1^+$ and B0 field maps a lower resolution of 7.5×5.6×6.2 mm per voxel was used. Other MPRAGE scan parameters include repetition time (TR) = 2200 ms, echo time (TE) = 2.77 ms, inversion time = 1050 ms, flip angle = 7°, bandwidth = 238 Hz/px, and scan time = 3:57 minutes. MPRAGE data were reconstructed as the root sum-of-squares of separate datasets acquired in CP-mode and CP2-mode, to improve the coverage into the neck of the structural information.

All data were acquired on a Siemens (Erlangen, Germany) Magnetom 7T scanner using a Nova Medical (Wilmington, MA) 8Tx/32Rx head coil under an institutional ethics agreement. To ensure consistency in the acquired B0 data, the tune-up B0 shim was used for all acquisitions. Data reconstruction and shim calculation were performed using MATLAB (The MathWorks, Natick, MA) on a system using an Intel (Intel, Santa Clara, CA) Xeon CPU E5-2680 (v4) running at 2.40 GHz with 14 cores and 28 logical processors.

Using the approach outlined in Section 2.1, large dynamic range $B_1^+$ field maps were measured and reconstructed for all 10 volunteers. For four subjects (Subjects 1, 2, 9, and 10), additional absolute $B_1^+$ maps were measured for validation purposes using two arbitrary RF shims (again acquired at reference voltages of 50 V, 100 V, and 175 V to facilitate validation with full spatial coverage).

### 2.3 Carotid artery RF shimming

In order to assess the theoretical upper limit for the boost in B1$^+$ in the neck that can be achieved using pTx RF shims, the total (theoretically) available B1$^+$ was evaluated *in vivo* on a voxel-by-voxel basis by summing the B1$^+$ magnitudes across the transmit channels.

For shim calculations and evaluation, hand-drawn vessel masks, comprising the internal carotid arteries (ICAs) and the Circle of Willis, were drawn for each subject from the MPRAGE images. These ROIs (Figure 4) were down-sampled to the resolution of the B1$^+$ and B0 data and used as masks for the RF shim calculations. Where needed, the ROIs were reduced to the areas corresponding to the carotid arteries.

Both phase-and-magnitude and phase-only RF shim combinations were calculated to assess any potential benefit of the extra degrees of freedom. Shims were calculated using cost functions that aim to maximize either the B1$^+$ magnitude, or the B1$^+$ homogeneity, or a combination of the magnitude and homogeneity. The cost function $\min \{< (1/B1^+)^2 >\}$ was used to maximize the magnitude, where a quadratic term is used to ensure simultaneous minimization of the required SAR to achieve a certain effective flip angle. The coefficient of variation (CoV) was used to maximize the B1$^+$ homogeneity: $\min \{CoV(B1^+)\} = \min \{std(B1^+)/< B1^+ >)\}$, where $std$ denotes the standard deviation. Finally, the combination of the magnitude and homogeneity was optimized using

$$\min \{CoV(B1^+) + \lambda * < (1/B1^+)^2 >\}, \qquad (1)$$

where $\lambda$ is a regularization parameter.

In order to assess the prospect of deploying a universal shim in the neck the convergence properties of universal neck RF shims were assessed using the data from the same 10 volunteers. To test the results when calculating a universal RF shim for *N* ($\leq 10$) subjects, a candidate RF shim was calculated for the first *N* subjects and its performance was evaluated using the multi-channel B1$^+$ data of all 10 subjects. The first *N* subjects were selected chronologically based on their acquisition date. For each comparison, tailored RF shims (where the shim is optimized for each subject separately) were included as an indication of the theoretical upper limit of the universal shim.

## 3. Results

### 3.1 Wide dynamic range multi-channel B1$^+$ mapping

Figure 2 shows an example of combining absolute B1$^+$ maps acquired using different transmit voltages and when using the exclusion criteria stated in Section 2.1. Figure 2b-d show the original CP-mode B1$^+$ maps acquired using transmit voltages of 50 V, 100 V, and 175 V, respectively, and Figure 2e shows the resulting combined B1$^+$ map. Figure 2f-h show masks of which voxel values were used to calculate the combined B1$^+$ map (after application of the three exclusion criteria), and Figure 2i shows how many values were used for each voxel to calculate the final combined B1$^+$ map.

Figure 3 shows an example MPRAGE image along with B1$^+$ maps from the first step in the B1$^+$ mapping pipeline. Figure 3b shows B0 off-resonance of up to -1.2 kHz in the neck, resulting in a B1$^+$ underestimation of up to 49% if not corrected. Figure 3c-e demonstrate the utility of B1TIAMO to increase the spatial coverage in areas with low CP-mode B1$^+$. Here, CP-mode and CP2-mode data were combined using a B1TIAMO weighting factor[19] (*m*) of 3. Figure 3e-f show the changes due to B0 correction of the absolute $B_1^+$ data based on the acquired B0 maps.

An overview of the data acquired from the 10 healthy volunteers is shown in Figure 4. For each subject a coronal slice of the MPRAGE data, a coronal projection of the hand-drawn vessel masks, and the reconstructed CP-mode $B_1^+$ map are shown. These data were acquired using a commonly used head pTx coil, so could also be useful for other research centres for the calculation of universal pTx shims or pulses, or for simulation purposes. Therefore, the multi-channel $B_1^+$ data and the B0 data are made openly available online (DOI: 10.5287/bodleian:ZB6Gk8QzN).

For Subjects 1, 2, 9, and 10, additional absolute $B_1^+$ maps were measured using two arbitrary RF shims. Figure 5 compares the measured (using the shim settings on the scanner) and predicted (combined multi-channel $B_1^+$ maps using the corresponding shim coefficients) $B_1^+$ maps for those validation shims. Visually good agreement is found for all four comparisons, with a root-mean-square error (RMSE) of 0.4 Hz/V in all cases and an average 95% confidence interval of ± 0.76 Hz/V.

### 3.2 Carotid artery RF shimming

The data from the 10 subjects shown in Figure 4 were used to study the potential $B_1^+$ benefits in the carotid arteries when using RF shims versus standard CP mode.

Figure 6 shows the CP-mode absolute $B_1^+$, the total (theoretically) available $B_1^+$, and the resultant CP-mode $B_1^+$ efficiency for two slices. Figure 6c shows that the theoretical upper limit of $B_1^+$ in the neck is (as expected) low compared to the central head region. In addition, Figure 6b and 6d show that CP-mode only utilizes 57 ± 13% of the theoretically available maximum $B_1^+$, resulting in an average $B_1^+$ magnitude in the neck for CP-mode of 2.5 ± 1.0 Hz/V. This suggests that utilising pTx should be able to improve this very low $B_1^+$ penetration, albeit never realising the theoretical maximum over a large region.

Figure 7 shows that a universal shim, generated by maximizing the average $B_1^+$ magnitude in the carotid arteries, is able to outperform CP-mode. However, it does not reach the theoretical voxel-by-voxel upper limit indicated by the available maximum $B_1^+$, which is unachievable using a standard RF shimming approach. Figure 7 also shows a large inferior-superior variation in the $B_1^+$ profile, since no coefficient of variation constraint was included in the calculation.

Figure 8 shows the neck shim performance within the vessel mask when a CoV optimization is included in the cost function using Equation 1. Both magnitude-and-phase, and phase-only $B_1^+$ shimming conditions were considered. Figure 8b shows that phase-only shimming performs almost as well as phase-and-magnitude shimming, with nearly identical results (differences < 1%) when using a regularization parameter $\lambda$ > 1.5. Based on the L-curve in Figure 8b, a regularization parameter, $\lambda$, of 1.7 is found to produce a reasonable trade-off between $B_1^+$ efficiency and minimizing the coefficient of variation.

When evaluating the number of subjects needed to form a universal shim (Figure 9), it is found that only a very small number of subjects is needed to achieve convergence. For all shim targets ($B_1^+$ magnitude optimized, CoV optimized, and optimized using Equation 1), universal shims perform only slightly worse than fully per-subject tailored shims, and substantially better than CP-mode. The final RF shim, calculated as a phase-only universal shim based on all 10 subjects and using $\lambda$ = 1.7, is shown in Figure 10. Over the full carotid artery masks, this universal RF shim achieves an average increase in $B_1^+$ magnitude of 37 ± 16% relative to CP-mode, whilst reducing the coefficient of variation by 22 ± 15%. When using tailored RF shims the corresponding improvements relative to CP-mode would be a $B_1^+$ magnitude increase of 42 ± 18% with a 27 ± 16% reduction in coefficient of variation.

# 4. Discussion

When combining $B_1^+$ maps acquired using CP and CP2 RF shims and transmit voltages using the proposed pipeline, a robust $B_1^+$ measurement can be obtained in the neck without compromising the $B_1^+$ accuracy in the head. Figure 4 shows that this increased coverage is consistently achieved, independent of subject size and position within the coil.

Figure 2 and Figure 3 show that both the combination of multiple transmit voltages and B1TIAMO contribute to increasing the coverage of the final $B_1^+$ into the neck. The inclusion masks in Figure 2f-h show that data reconstructed from low-voltage acquisitions are mainly used for the high-$B_1^+$ areas in the center of the brain and close to the transmit elements, while high-voltage data mainly contribute accurate information in the neck. These observations are consistent with the assumptions that motivated the use of multiple transmit voltages. Furthermore, they indicate that the exclusion criteria in Section 2.1, which do not make use of the spatial location of voxels, can accurately determine which transmit values to include at different spatial locations.

Despite the increased effective dynamic range when using multiple transmit voltages, even with a single shim (e.g. CP-mode) no accurate $B_1^+$ information can be acquired in locations which have very low $B_1^+$ values due to destructive interference of the transmit fields. In such cases, Figure 3 confirms that including CP2-mode and B1TIAMO combination of CP and CP2 modes provides complementary information and therefore yields improved spatial extent of the $B_1^+$ maps.

Although combining data acquired using different RF shims and transmit voltages provides $B_1^+$ information with a larger spatial extent, B0 off-resonance effects can reduce the accuracy of the measured $B_1^+$ values. In this work, we used rectangular pulses that required an additional B0 correction step (Figure 3) to obtain accurate values in areas with high B0 off-resonance (up to -1.2 kHz were observed). Alternatively, broadband full-passage hyperbolic secant (HS8) pulses[22] can be used for the pre-saturated TurboFLASH acquisitions[17] to reduce the B0-dependence of the $B_1^+$ estimates.

When comparing predicted $B_1^+$ maps (reconstructed using the proposed pipeline) and measured $B_1^+$ maps (acquired directly on the scanner) for arbitrary RF shims (Figure 5), excellent agreement can be observed throughout the imaging volume. The RMSE values of around 0.4 Hz/V indicate that some differences remain between the predicted and acquired maps. However, some of this remaining disagreement may be caused by inaccuracies in the measured $B_1^+$ maps rather than the predicted $B_1^+$ maps. For example, in the low $B_1^+$ regions in the middle of the head for Validation Shim 2, discontinuities which do not typically appear in $B_1^+$ maps are visible in the measured $B_1^+$ maps, while the predicted $B_1^+$ maps remain spatially smooth. This is also visible in the scatter plots in Figure 5, where higher measured values are observed for voxels with low predicted $B_1^+$ values. Other voxels with larger errors (visible in the third row in Figure 5) are located at air-tissue interfaces, indicating that partial volume effects might be a further source of the remaining errors.

The proposed multi-channel $B_1^+$ mapping method requires a total of 10 separate acquisitions (6 absolute $B_1^+$ maps, 3 sets of relative $B_1^+$ maps, and 1 B0 map) with a total scan time of 6:45 minutes for the reconstruction of a single set of multi-channel $B_1^+$ maps. This additional scan time would be a limiting factor if acquiring subject-specific field maps at the start of a clinical exam. However, this is no longer a limitation if using the method to acquire a $B_1^+$ database for the calculation of universal RF shims or pulses.

Figure 6 shows that the $B_1^+$ efficiency of CP-mode is low in the neck (57 ± 13% along the carotid arteries), meaning that there is substantial opportunity for improvement using (universal) RF shims. The results in Figure 7, which show the performance of a universal shim aiming to minimize the required RF power along the carotid arteries, confirm that a universal shim can substantially improve the carotid $B_1^+$, while showing a reduction in $B_1^+$ in the Circle of Willis (which, depending on the application, may be an advantage or a disadvantage). However, it also evident that there is substantial $B_1^+$ variation along the vessel, suggesting that the shim performance can be improved by adding a $B_1^+$ homogeneity constraint.

The results in Figure 8 show that combining such a $B_1^+$ homogeneity optimization with a $B_1^+$ magnitude optimization (using Equation 1) can improve the average homogeneity with a minimal reduction in $B_1^+$ magnitude. When using a regularization parameter, $\lambda$, of 1.7, the CoV is reduced by 12% while the average $B_1^+$ magnitude is reduced by only 5%. Figure 8b indicates that the results using phase-only shims are nearly equal to those of phase-and-magnitude shims when using $\lambda > 1.5$, indicating an inherent requirement for high $B_1^+$ utilization from all channels in order to achieve sufficient $B_1^+$ in the neck.

The universal shim convergence comparison in Figure 9 shows that a universal shim in the vessels in the neck can easily be found: good results are already found when universal shims are calculated for a single subject, and (when using $\lambda = 1.7$) no further improvement is observed when including more than 4 subjects in the shim calculation. Furthermore, Figure 9 shows that universal RF shims perform almost as effectively as fully tailored per-subject RF shims, while consistently outperforming CP-mode in terms of both $B_1^+$ magnitude and coefficient of variation.

The results in Figure 10 show that, using a universal RF shim and $\lambda = 1.7$, the $B_1^+$ magnitude in the vessels in the neck can be increased by 37%, whilst reducing the coefficient of variation by 22%. This can be achieved using phase-only RF shimming and does not require phase-and-magnitude RF shimming. Since a linear increase in $B_1^+$ for phase-only RF shimming implies a quadratic reduction in required SAR to achieve a given flip angle, the reported 37% increase in $B_1^+$ magnitude corresponds to a 47% reduction in SAR. These results are based on optimization of the $B_1^+$ over the entire region of the carotid vessels in the vessel masks in Figure 4. For some applications, in particular for ASL, excitation targets can consist of a smaller portion of these vessels, for example when only labelling in a certain plane or when using vessel-selective ASL[23]. In such cases, the optimization is less constrained, allowing for larger improvements in RF shim performance. For example, when only including the left internal carotid artery as a shim target, a phase-only universal RF shim can simultaneously achieve a 44% increase in $B_1^+$ magnitude and a 41% decrease in CoV relative to CP mode (data not shown).

It should be noted that the vessel masks used in this study were drawn based on the vasculature of healthy volunteers. Although the results presented here indicate consistently improved $B_1^+$ in the carotid arteries for subjects with typical (vascular) anatomy, both the $B_1^+$ fields and the locations of the vessels in the neck might be different for patients with non-standard anatomies. Figure 10e shows a consistent increase in $B_1^+$ across both the left and the right side of the neck when using the proposed shim, with a decrease in $B_1^+$ in the centre of the neck. Although morphological variations in the shape of the internal carotid arteries increases with age[24], a patient study into the variability of the medial location of the ICAs[25] found that the ICAs of most (96.1%) patients are located within the lateral half on each side of the neck and would therefore be expected to achieve substantial $B_1^+$ improvements even for the universal RF shim. 3.6% of the ICAs were found in the medial half of the lateral mass, roughly corresponding to the transition areas (black) in Figure 10e. The $B_1^+$ reduction in Figure 10e would only correspond to the location of the ICAs in the remaining 0.3% of patients, who had ICAs located medial to the lateral mass.

Furthermore, the results presented here are all based on simple RF shims which are constrained to the superposition patterns that can be achieved using the available transmit channels. Since the average $B_1^+$ efficiency of the proposed universal RF shim is 74 $\pm$ 3%, it is expected that further improvements can be achieved when using more advanced dynamic pTx pulses, where additional degrees of freedom are introduced by continuously changing the pTx coefficients in combination with the gradient waveforms and pulse amplitudes. This can be used to further achieve improved $B_1^+$ homogeneity and/or localization. However, an advantage of using RF shims is that they can directly be implemented into existing sequences without requiring further pulse design or introducing sequence timing restrictions, thereby not increasing experimental complexity for existing sequences while still achieving substantially improved

$B_1^+$ efficiency and homogeneity.

# 5. Conclusion

Combining $B_1^+$ data acquired using different voltages with CP- and CP2-mode RF shims allows the reconstruction of accurate multi-channel head-and-neck $B_1^+$ maps for pTx head coils at 7T. Using this, universal RF shims can be designed that increase the $B_1^+$ magnitude in the arteries in the neck by 37%, while also improving the homogeneity. This is possible using phase-only universal RF shims, facilitating easy implementation in existing sequences at 7T.

**Acknowledgements**

The Wellcome Centre for Integrative Neuroimaging is supported by core funding from the Wellcome Trust (203139/Z/16/Z). We also thank the Dunhill Medical Trust and the NIHR Oxford Biomedical Research Centre for support (PJ). MdB acknowledges studentship support from Siemens Healthineers and the Dunhill Medical Trust. AH acknowledges support from the BHF Centre of Research Excellence, Oxford (RE/13/1/30181).

**Data availability statement**

Both the Matlab code for the wide dynamic range multi-channel B1+ reconstruction method ([git.fmrib.ox.ac.uk/ndcn0873/b1_pipeline_reconstruction](git.fmrib.ox.ac.uk/ndcn0873/b1_pipeline_reconstruction)) and our 10-subject multi-channel $B_1^+$ database (DOI: 10.5287/bodleian:ZB6Gk8QzN) acquired using that method are openly available online. The online database also includes the corresponding B0 maps. In line with GDPR requirements, higher resolution structural MPRAGE data are available on request.

# Figures

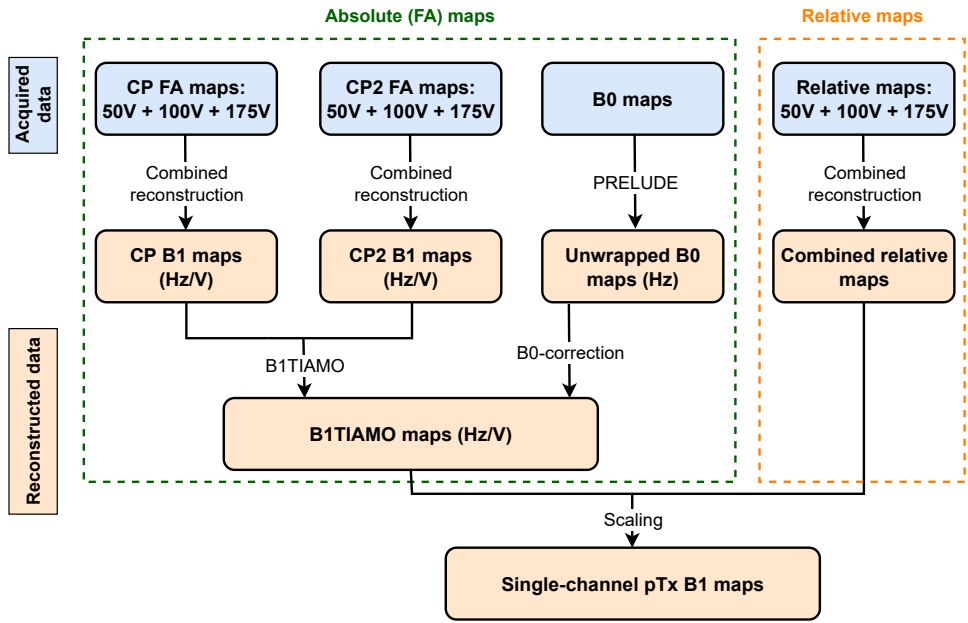

*Figure 1:* Schematic of the proposed $B_1^+$ map acquisition and processing pipeline. In total, 10 different datasets are acquired (blue: 6 FA maps, 1 B0 map, and 3 sets of relative maps) to reconstruct a single set of multi-channel $B_1^+$ maps.

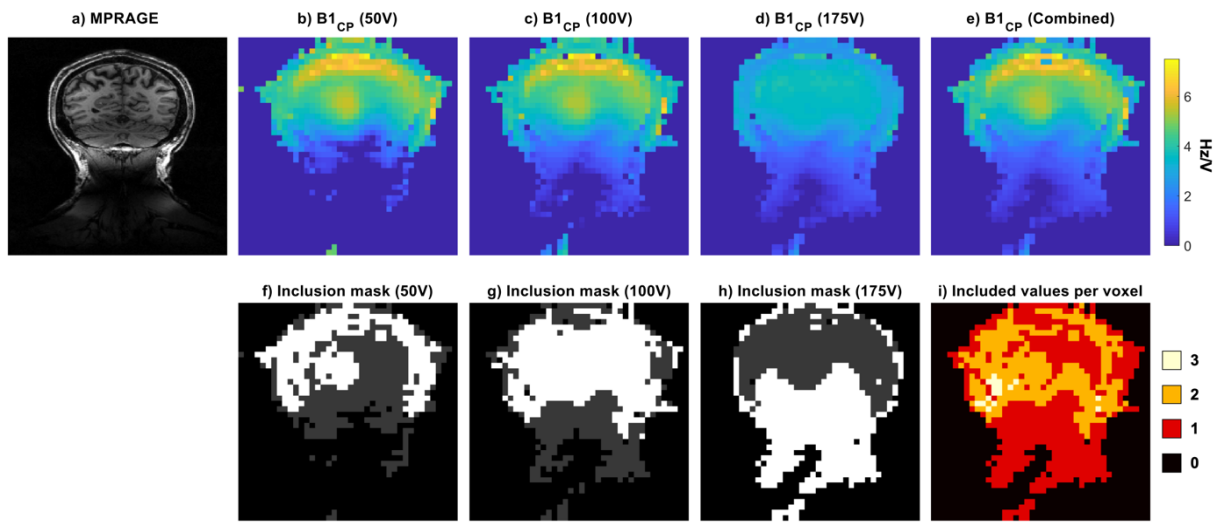

*Figure 2:* Absolute $B_1^+$ maps acquired at different voltages *(b-d)* are combined to obtain a single map with an increased combined dynamic range *(e)*. Voxel values from individual scans were included or excluded based on the signal levels in the reference images and exclusion criteria to impose consistency in the acquired values relative to the other acquired datasets, resulting in inclusion masks for each dataset *(f-h)*. Figure *(i)* shows the total number of included values for each voxel in the slice.

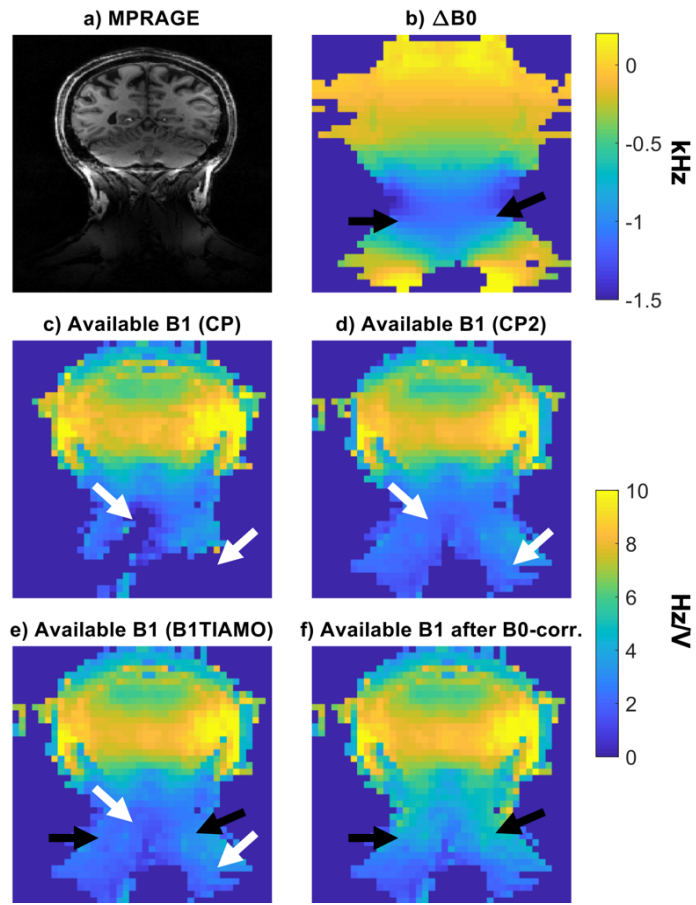

*Figure 3:* The use of B1TIAMO to mitigate signal loss in regions with low native $B_1^+$ in CP-mode to obtain improved spatial coverage *(c-e)*, and the consecutive correction for $B_1^+$ underestimation in the presence of high B0 inhomogeneity *(f)* based on measured B0 off-resonance fields *(b)*. White arrows indicate examples of improved spatial coverage due to B1TIAMO; black arrows indicate areas with substantial B0 offsets (up to -1.2 kHz), resulting in substantial $B_1^+$ underestimation if no B0 correction is applied. All $B_1^+$ data are shown in terms of the available $B_1^+$.

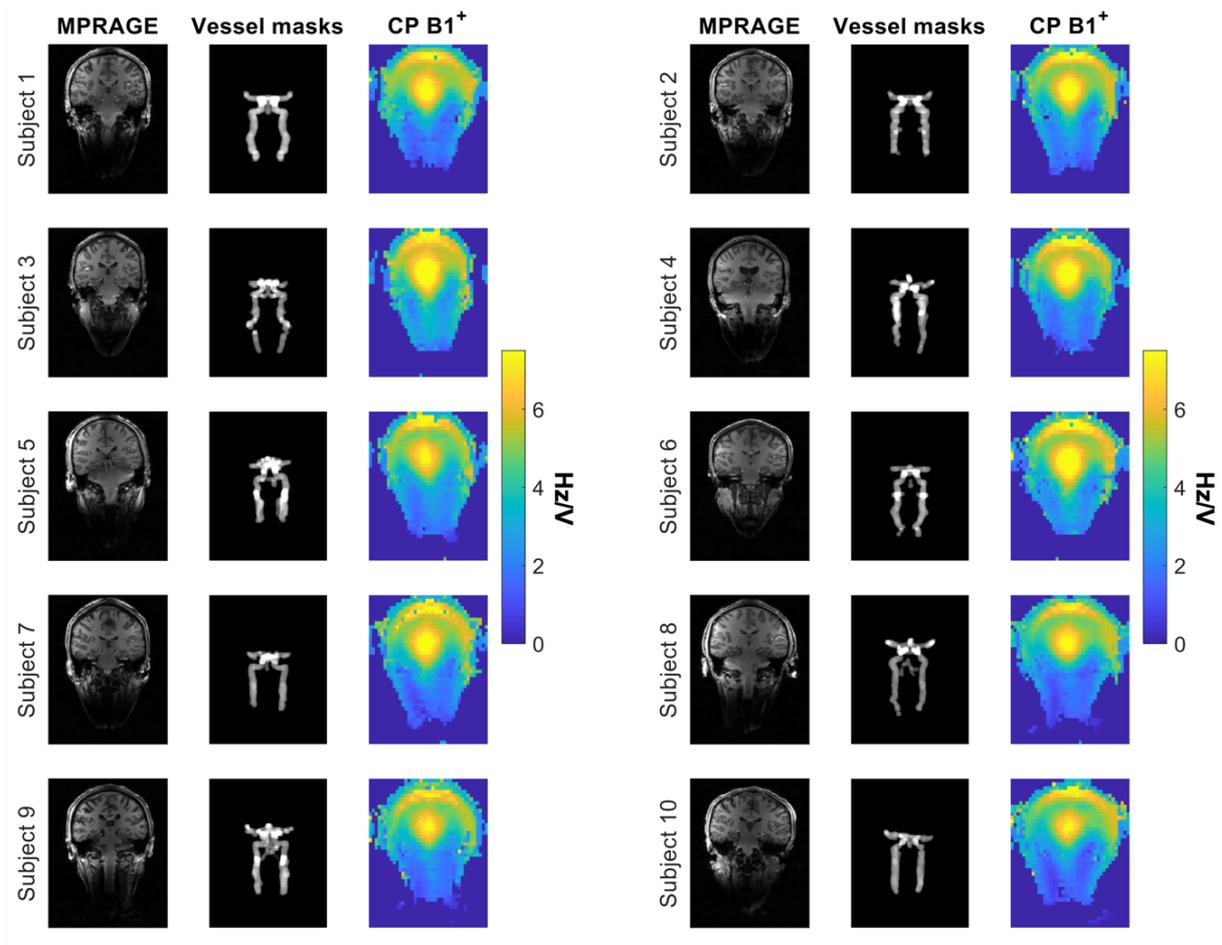

*Figure 4:* Central coronal slices of the 10-subject database (DOI: 10.5287/bodleian:ZB6Gk8QzN) that was acquired using the proposed method. All data were acquired using the same FOV in coil coordinates. Columns show MPRAGE data (root-sum-of-squares of CP-mode and CP2-mode to improve structural visibility in the neck), left column; coronal projections of the arterial vessel masks corresponding to the MPRAGE data, middle column; and the CP-mode $B_1^+$ map for each subject, right column. The CP $B_1^+$ maps shown here are synthetic maps generated from the multi-channel $B_1^+$ data.

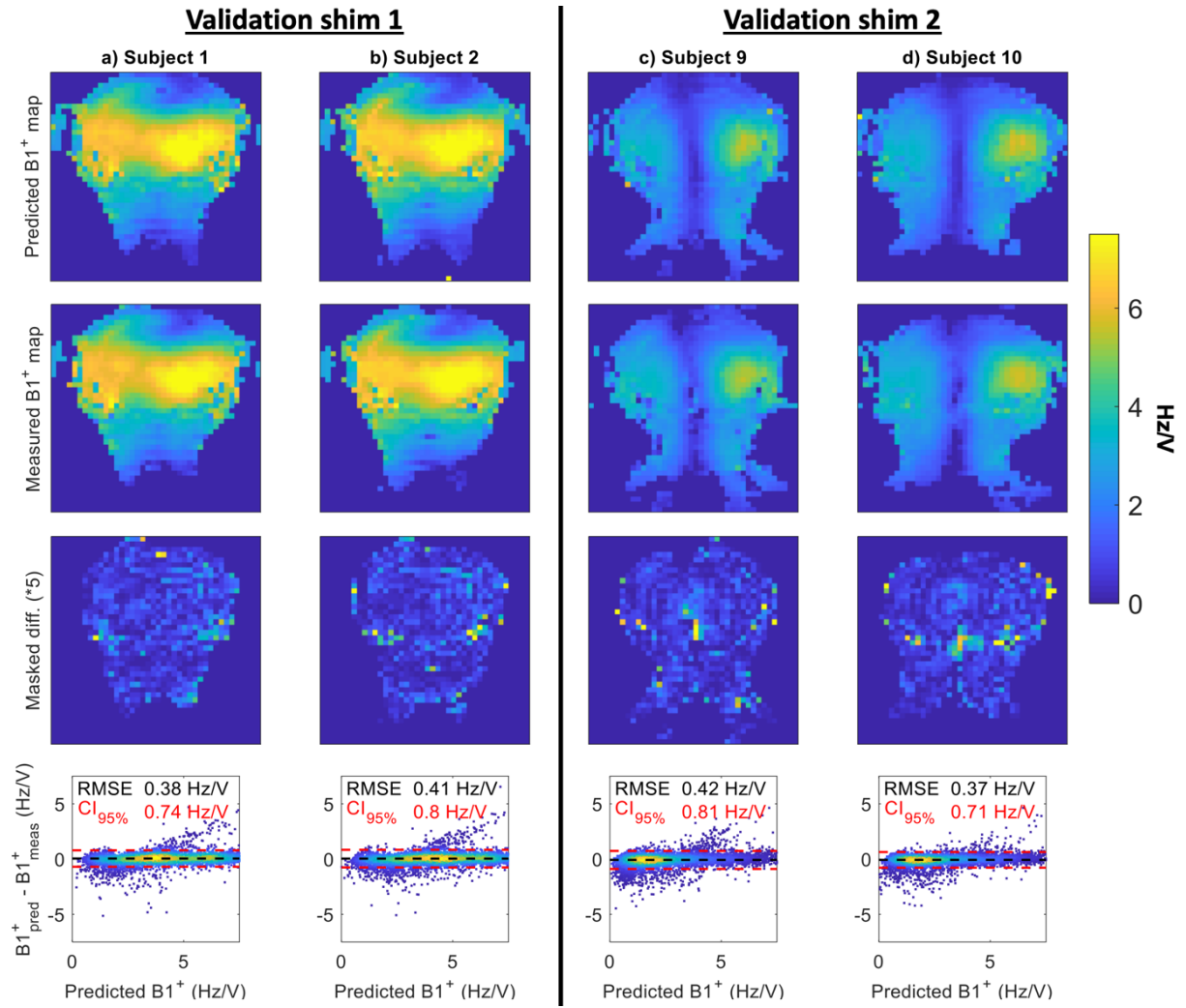

*Figure 5: Evaluation of the agreement between predicted (1st row; calculated from reconstructed multi-channel $B1^+$ maps) and measured (2nd row; acquired on the scanner using the same shim coefficients) $B1^+$ magnitude maps for two arbitrary RF shims. The 3rd and 4th rows show the absolute difference (using a 5-fold boosted colour scale) between the images in the first two rows, and the difference between the predicted ($B1^+_{pred}$) and measured ($B1^+_{meas}$) voxel-wise values within the overlapping region. Dashed black lines indicate the mean errors, with dashed red lines indicating the means $\pm 95\%$. Printed values indicate the root-mean-square errors (RMSE, black) and 95% confidence intervals ($CI_{95\%}$, red).*

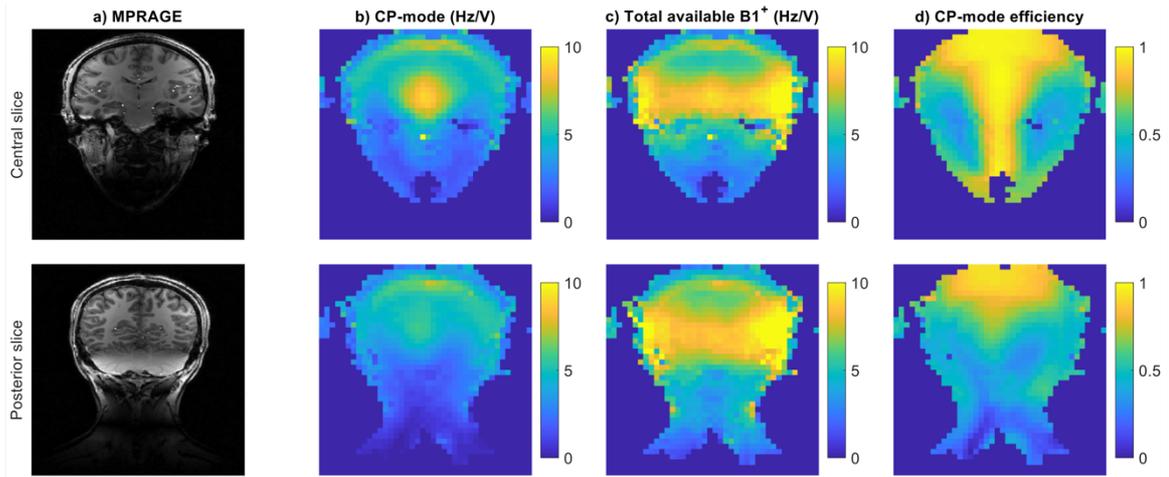

*Figure 6:* Two coronal slices from an example subject (Subject 1), showing a central slice (top row) and a more posterior slice (bottom row). Column **(b)** shows the $B1^+$ map (Hz/V) in CP mode. Column **(c)** shows the maximum possible $B1^+$ for each voxel (calculated as the sum of magnitude $B1^+$ per channel). Column **(d)** shows the CP mode efficiency, calculated as the ratio of CP mode $B1^+$ divided by total available $B1^+$, indicating the loss of potential $B1^+$ arising when using CP mode.

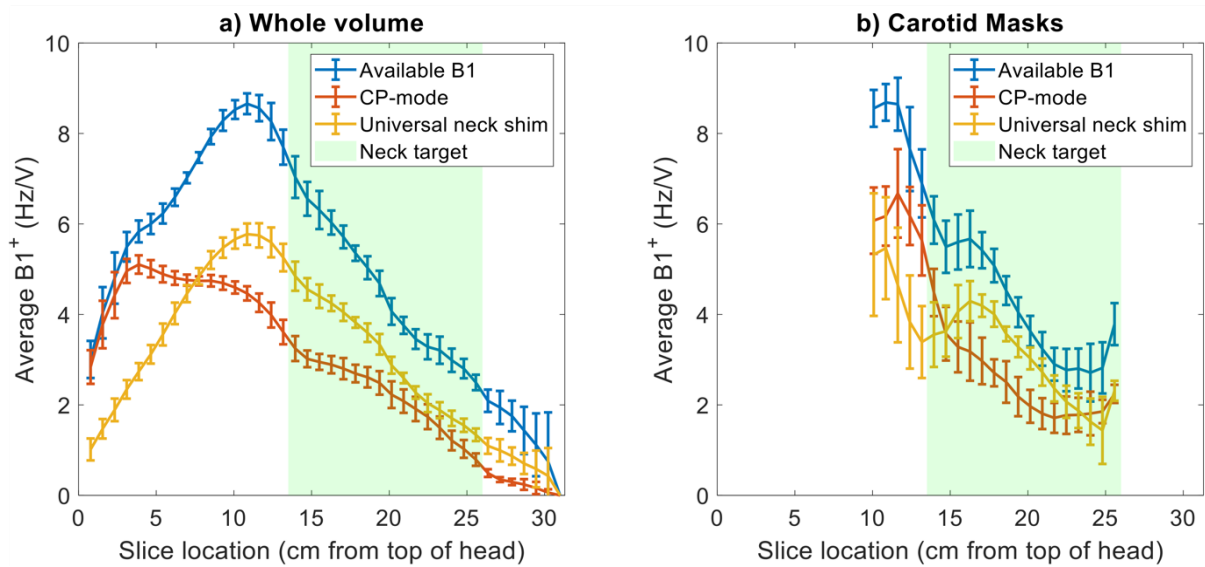

*Figure 7*: Plots showing the average $B1^+$ across all 10 subjects for CP mode (orange lines), for the total available $B1^+$ (blue lines), and the $B1^+$ achieved using a universal neck shim (yellow lines; calculated using the $B1^+$ magnitude cost function $min \{< (1/B1^+)^2 >\}$). **(a)** shows the results averaged over the whole head volume (with the neck region indicated) and **(b)** shows the data within the vessel masks only. Note that the $B1^+$ superior to the neck mask is reduced for the universal shim relative to CP mode, whereas the $B1^+$ within the neck mask is increased for the universal shim relative to CP mode.

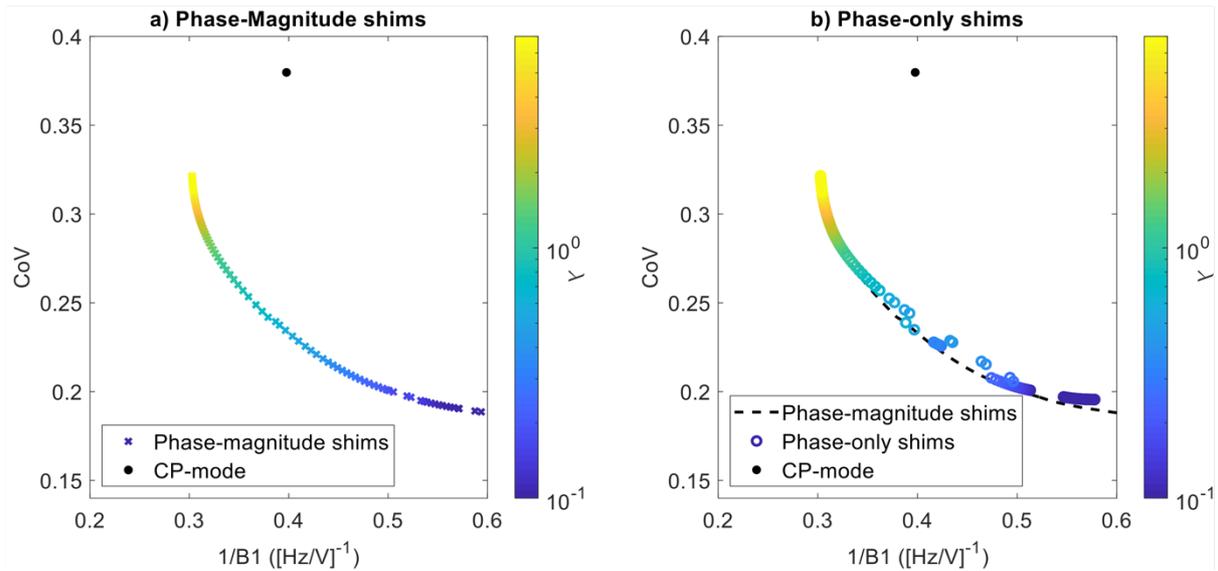

*Figure 8:* Plots investigating the trade-off between the coefficient of variation (CoV) of $B_1^+$ within the vessel mask versus the achieved $B_1^+$ magnitude (expressed as $1/B_1^+$). The desired regularization value, $\lambda$, that reduces CoV whilst retaining a strong $B_1^+$ is found at approximately $\lambda=1.7$ (green region). **(a)** shows universal shims that allow both phase and magnitude to change per channel. **(b)** shows universal shims that allow only phase to change per channel. The dashed black line in (b) shows the data from (a) overlaid as a guide to the eye. The black dot indicates CP mode.

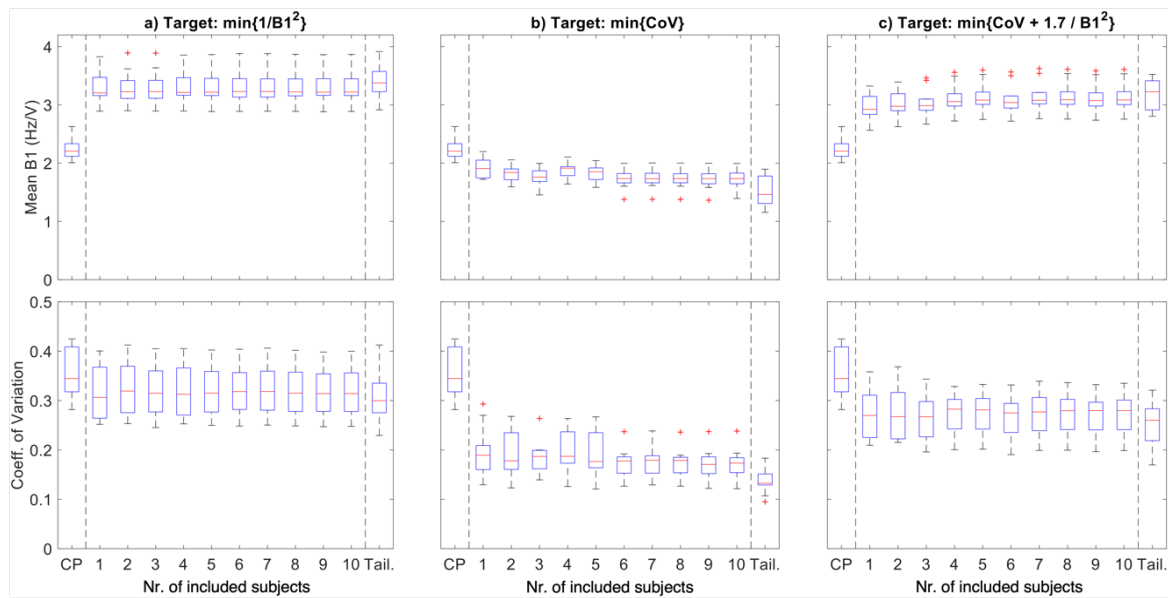

*Figure 9:* Plots showing the number of subjects needed to generate a universal neck shim (based on phase-only RF shims). For each plot the CP-mode mean $B_1^+$ and CoV are shown for reference, followed by the relevant metrics for universal neck shims generated from increasing numbers of subjects. The final column for each plot shows the result when per-subject tailored shims are used (denoted Tail.). The different columns show the results for three different cost functions: **(a)** minimizing $\{1/B_1^2\}$; **(b)** minimizing $\{CoV\}$; and **(c)** minimizing the optimum combination of $\{1/B_1^2\}$ and $\{CoV\}$ with regularization value $\lambda = 1.7$.

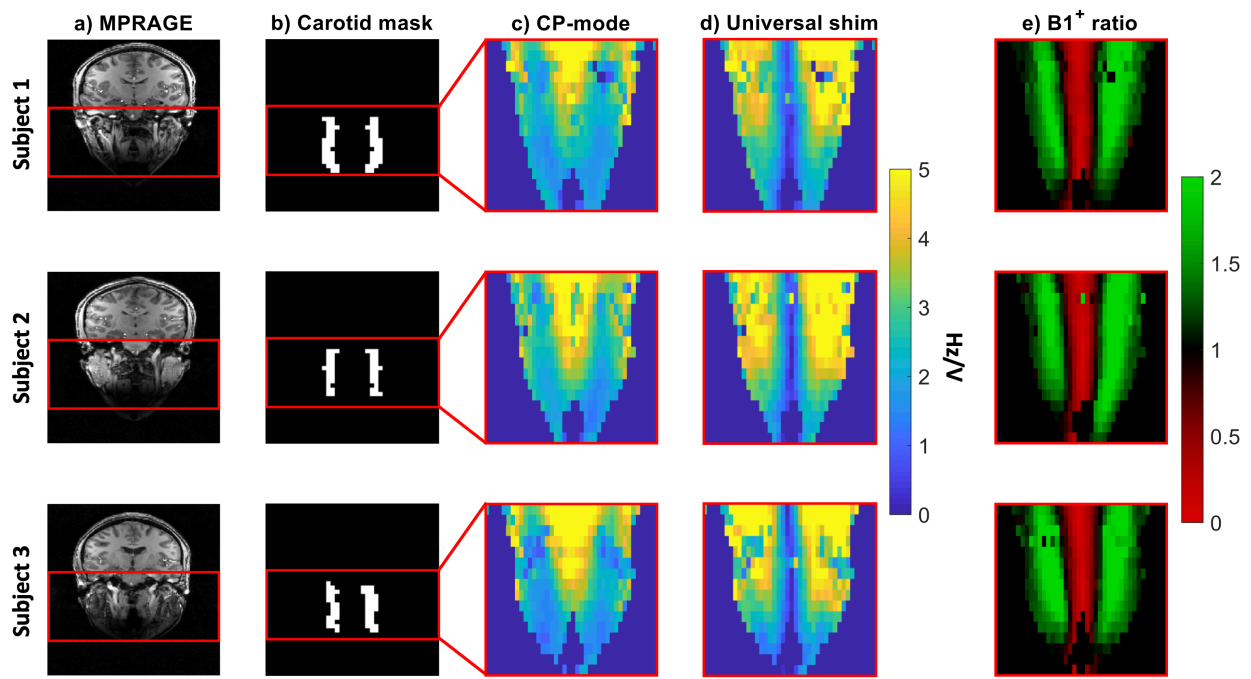

*Figure 10: An example RF shim for the carotid arteries, shown for Subjects 1-3. Columns show **(a)** an MPRAGE slice containing a superior segment of the internal carotid arteries; **(b)** the corresponding portion of the down-sampled vessel masks; **(c)** the CP-mode $B_1^+$ field in the vessel region (shown for the area corresponding to the red box in (a-b) to allow for a limited colour bar range); **(d)** the corresponding $B_1^+$ field when using a phase-only universal shim calculated using regularization $\lambda = 1.7$; and **(e)** the ratio between (d) and (c), showing where the $B_1^+$ increases (green) or decreases (red) when using the universal RF shim instead of CP-mode.*